\documentclass[english,aps,superscriptaddress]{revtex4}
\usepackage[T1]{fontenc}
\usepackage[latin9]{inputenc}
\usepackage{verbatim}
\usepackage{amsmath}
\usepackage{graphicx}

\makeatletter
\@ifundefined{textcolor}{}
{%
 \definecolor{BLACK}{gray}{0}
 \definecolor{WHITE}{gray}{1}
 \definecolor{RED}{rgb}{1,0,0}
 \definecolor{GREEN}{rgb}{0,1,0}
 \definecolor{BLUE}{rgb}{0,0,1}
 \definecolor{CYAN}{cmyk}{1,0,0,0}
 \definecolor{MAGENTA}{cmyk}{0,1,0,0}
 \definecolor{YELLOW}{cmyk}{0,0,1,0}
}

\makeatother

\usepackage{babel}
\begin{document}
\title{Quark spin correlation inside hyperons}
\author{Lucia Oliva}
\email{lucia.oliva@ct.infn.it}

\affiliation{INFN-Sezione di Catania, Via Santa Sofia 64, I-95123 Catania, Italy}
\affiliation{Department of Physics and Astronomy ``Ettore Majorana'', University
of Catania, Via Santa Sofia 64, I-95123 Catania, Italy}
\author{Qun Wang}
\email{qunwang@ustc.edu.cn}

\affiliation{Department of Modern Physics and Anhui Center for Fundamental Sciences
in Theoretical Physics, University of Science and Technology of China,
Hefei, Anhui 230026, China}
\affiliation{School of Mechanics and Physics, Anhui University of Science and Technology,
Huainan, Anhui 232001, China}
\author{Xin-Nian Wang}
\email{xnwang@ccnu.edu.cn}

\affiliation{Central China Center for Nuclear Theory and Institute of Particle
Physics, Central China Normal University, Wuhan 430079, China}
\affiliation{Institut fuer Theoretische Physik, Johann Wolfgang Goethe Universitaet,
Max-von-Laue-Str. 1, D-60438 Frankfurt am Main, Germany}
\begin{abstract}
The global spin polarization of hyperons in heavy-ion collisions have
been investigated by including spin correlation effects among their
constituent quarks. The available data on global spin polarizations
of hyperons and spin alignments of vector mesons provide constraints
on phase space functions of the spin polarization and correlation.
These constraints can lead to inequalities under some approximations,
which might provide possible clues for the presence of quark spin
correlation inside hyperons at lower collision energies.
\end{abstract}
\maketitle

\section{Introduction}

Spin physics in heavy-ion collisions has attracted a lot of interest
since the first measurement of the global polarization of $\Lambda$
hyperons by the STAR Collaboration at the Relativistic Heavy Ion Collider
(RHIC) \citep{STAR:2017ckg} followed by a series of later experiments
at different energies or under different conditions \citep{STAR:2018gyt,STAR:2020xbm,STAR:2021beb,ALICE:2019onw,HADES:2022enx}.
This confirms the theoretical predictions made almost two decades
ago \citep{Liang:2004ph,Liang:2004xn,Voloshin:2004ha,Betz:2007kg,Gao:2007bc,Becattini:2007sr}.

Recently the STAR Collaboration measured nonvanishing global spin
alignment of the $\phi$ vector meson \citep{STAR:2022fan}, implying
a strong spin correlation between the strange and antistrange quark
that combine to form the $\phi$ meson \citep{Sheng:2019kmk,Sheng:2022wsy,Sheng:2022ffb},
for recent reviews, see, e.g. Refs. \citep{Chen:2023hnb,Chen:2024afy}.
The experimental observations and corresponding theoretical studies
provide an opportunity to study the spin correlations \citep{Pang:2016igs,Sheng:2025puj}
in high energy heavy-ion collisions for the first time. Very recently
the STAR Collaboration also measured the spin correlation between
$\Lambda$ and $\bar{\Lambda}$ in proton-proton collisions \citep{STAR:2025njp}
which are inherited from spin-correlated strange quark-antiquark virtual
pairs \citep{Tornqvist:1980af,Tornqvist:1986pe,Ellis:2011kq,Gong:2021bcp,Wu:2024asu},
which could serve as probes into QCD confinement.  

A systematic description of quark (antiquark) spin correlations in
quark matter can be provided by the multi-quark spin density matrix
(SDM) in spin space \citep{Lv:2024uev,Zhang:2024hyq}. From the multi-quark
SDM, one can obtain the SDM for a vector meson/hyperon through the
transition operator to convert quark-antiquark/three quarks to the
vector meson/hyperon. The hadron's spin polarization can then be determined
by its SDM and depends not only on those of its constituent quarks
but also on the spin correlations among them. The spin correlations
in previous calculations of global and local spin polarizations for
hyperons were all neglected: it seems that the data could be well
described without the need for spin correlations \citep{Xia:2018tes,Karpenko:2016jyx,Sun:2017xhx,Li:2017slc,Wei:2018zfb,Vitiuk:2019rfv,Becattini:2020ngo,Fu:2020oxj,Ryu:2021lnx,Li:2021zwq,Palermo:2024tza,Sapna:2025yss}.
However, there is an indication for a possible disagreement between
the data and model calculations for the global polarization of $\Omega$
hyperons \citep{zhiwan_xu:qm2025}. A natural question arises if the
possible disagreement has to do with neglecting the spin correlation
among strange quarks in $\Omega$. In this paper, we try to examine
this problem by including the spin correlation effects into the spin
polarizations of hyperons. Based on density operators for quarks (antiquarks)
in spin-momentum space, we can derive spin Wigner functions as the
SDM elements in phase space for vector mesons and baryons which are
expressed as phase space integrals of their constituent quarks' SDM
weighted by their coordinate wave functions, constituent quark distributions
and Clebsch-Gordan coefficients. In this way we obtain spin polarizations
of vector mesons and hyperons depending on spin polarization and correlation
functions for quarks. In comparison with data for global spin polarizations
of hyperons and spin alignments of vector mesons, one can derive constraints
for quark spin correlations under some approximations.

By convention, the components of a three-vector can be expressed as
$\boldsymbol{a}=(a_{x},a_{y},a_{z})=(a^{x},a^{y},a^{z})$, where we
do not distinguish superscripts and subscripts for its spatial components. 


\section{Spin Wigner functions for vector mesons and baryons}

Relativistic kinetic theory in the Wigner-function formalism \citep{Heinz:1983nx,Elze:1986hq,Vasak:1987um,Shin:1992nj,Bialynicki-Birula:1993wgw,Zhuang:1995pd,Zhuang:1998bqx,Blaizot:2001nr}
can describe the spin polarization of spin-1/2 massless \citep{Son:2012wh,Son:2012zy,Stephanov:2012ki,Gao:2012ix,Chen:2012ca,Hidaka:2016yjf,Gao:2017gfq,Gao:2018wmr,Huang:2018wdl,Carignano:2018gqt,Liu:2018xip,Gao:2019zhk,Yang:2020mtz,Hou:2020mqp,Manuel:2021oah}
and massive fermions \citep{Fang:2016vpj,Weickgenannt:2019dks,Gao:2019znl,Hattori:2019ahi,Weickgenannt:2020aaf,Liu:2020flb,Weickgenannt:2021cuo,Sheng:2021kfc,Sheng:2022ssd}
determined by the axial-vector or tensor component of the Wigner function,
see, e.g., Refs. \citep{Hidaka:2022dmn,Becattini:2024uha} for recent
reviews on this topic.

The spin Wigner functions (SWF) are actually the spin density matrix
elements in phase space. The SWF for vector mesons and baryons can
be obtained from those for quark-antiquark and three quarks through
the spin-flavor or SU(6) wave functions \citep{Yang:2017sdk,Sheng:2019kmk,Lv:2024uev,Zhang:2024hyq}.
In this section, we will derive the SWF in phase space for vector
mesons and baryons in non-relativistic quantum theory. 

\subsection{Spin density operators for one-particle, two-particle and three-particle
systems}

We start from the density operator for a non-relativistic spin-1/2
particle on mass-shell defined as 
\begin{eqnarray}
\rho_{(1)}^{\mathrm{op}} & = & \sum_{r,s}\int[d^{3}\boldsymbol{p}_{1}][d^{3}\boldsymbol{p}_{2}]W(r,s;\boldsymbol{p}_{1},\boldsymbol{p}_{2})\left|r,\boldsymbol{p}_{1}\right\rangle \left\langle s,\boldsymbol{p}_{2}\right|\nonumber \\
 & = & \sum_{r,s}\int[d^{3}\boldsymbol{p}][d^{3}\boldsymbol{q}]d^{3}\boldsymbol{x}e^{-i\boldsymbol{q}\cdot\boldsymbol{x}}W(r,s;\boldsymbol{x},\boldsymbol{p})\left|r,\boldsymbol{p}+\frac{\boldsymbol{q}}{2}\right\rangle \left\langle s,\boldsymbol{p}-\frac{\boldsymbol{q}}{2}\right|\nonumber \\
 & = & \sum_{r,s}\int d^{3}\boldsymbol{x}\int[d^{3}\boldsymbol{p}]W(r,s;\boldsymbol{x},\boldsymbol{p})\int[d^{3}\boldsymbol{q}]e^{-i\boldsymbol{q}\cdot\boldsymbol{x}}\left|r,\boldsymbol{p}+\frac{\boldsymbol{q}}{2}\right\rangle \left\langle s,\boldsymbol{p}-\frac{\boldsymbol{q}}{2}\right|.\label{eq:density-op}
\end{eqnarray}
Here $r,s=\pm1/2$ denote the spin states along the spin quantization
direction. We have used the shorthand notation $[d^{3}\boldsymbol{q}]\equiv d^{3}\boldsymbol{q}/(2\pi)^{3}$,
and changed in second line the momenta $(\boldsymbol{p}_{1},\boldsymbol{p}_{2})$
to $(\boldsymbol{p},\boldsymbol{q})$ with $\boldsymbol{p}_{1,2}=\boldsymbol{p}\pm\boldsymbol{q}/2$,
we also defined the Wigner function $W(r,s;\boldsymbol{x},\boldsymbol{p})$
as 
\begin{align}
W(r,s;\boldsymbol{q},\boldsymbol{p})\equiv & W(r,s;\boldsymbol{p}+\boldsymbol{q}/2,\boldsymbol{p}-\boldsymbol{q}/2)=\int d^{3}\boldsymbol{x}e^{-i\boldsymbol{q}\cdot\boldsymbol{x}}W(r,s;\boldsymbol{x},\boldsymbol{p}),\nonumber \\
W(r,s;\boldsymbol{x},\boldsymbol{p})= & \int[d^{3}\boldsymbol{q}]e^{i\boldsymbol{q}\cdot\boldsymbol{x}}W(r,s;\boldsymbol{q},\boldsymbol{p})\nonumber \\
= & \int[d^{3}\boldsymbol{q}]e^{i\boldsymbol{q}\cdot\boldsymbol{x}}\left\langle r,\boldsymbol{p}+\frac{\boldsymbol{q}}{2}\left|\rho_{(1)}^{\mathrm{op}}\right|s,\boldsymbol{p}-\frac{\boldsymbol{q}}{2}\right\rangle ,
\end{align}
where $\boldsymbol{x}$ is the particle's position conjugate to the
relative momentum $\boldsymbol{q}$, and we have used the orthogonality
relation for momentum states 
\begin{equation}
\left\langle r,\boldsymbol{p}\right.\left|s,\boldsymbol{k}\right\rangle =(2\pi)^{3}\delta_{rs}\delta^{(3)}(\boldsymbol{p}-\boldsymbol{k}).
\end{equation}
We assume that the Wigner function of a non-relativistic particle
can be factorized into the polarized and unpolarized parts
\begin{align}
W(r,s;\boldsymbol{x},\boldsymbol{p})= & f(\boldsymbol{x},\boldsymbol{p})\left\langle r\right|\hat{\rho}_{(1)}(\boldsymbol{x},\boldsymbol{p})\left|s\right\rangle \nonumber \\
\equiv & f(\boldsymbol{x},\boldsymbol{p})\frac{1}{2}\left[\delta_{rs}+\boldsymbol{\sigma}_{rs}\cdot\boldsymbol{P}(\boldsymbol{x},\boldsymbol{p})\right]\label{eq:one-particle-rho}
\end{align}
where $\boldsymbol{\sigma}=(\sigma_{x},\sigma_{y},\sigma_{z})$ denote
the Pauli matrices, $f(\boldsymbol{x},\boldsymbol{p})$ and $\boldsymbol{P}(\boldsymbol{x},\boldsymbol{p})$
are the unpolarized and polarized distribution functions respectively,
and the second line defines the SDM for one particle $\hat{\rho}_{(1)}(\boldsymbol{x},\boldsymbol{p})$. 

The density operator for two spin-1/2 particles can be defined as
\begin{align}
\rho_{(12)}^{\mathrm{op}}= & \sum_{r_{1},s_{1},r_{2},s_{2}}\int\prod_{n=1,2}[d^{3}\boldsymbol{p}_{n}][d^{3}\boldsymbol{q}_{n}]d^{3}\boldsymbol{x}_{n}e^{-i\boldsymbol{q}_{n}\cdot\boldsymbol{x}_{n}}\nonumber \\
 & \times W\left(\{r_{n},s_{n};\boldsymbol{x}_{n},\boldsymbol{p}_{n}\},n=1,2\right)\prod_{n=1,2}^{\otimes}\left|r_{n},\boldsymbol{p}_{n}+\frac{\boldsymbol{q}_{n}}{2}\right\rangle \left\langle s_{n},\boldsymbol{p}_{n}-\frac{\boldsymbol{q}_{n}}{2}\right|,\label{eq:rho-12}
\end{align}
where the joint Wigner function for two particles can be put into
the form 
\begin{equation}
W\left(\{r_{n},s_{n};\boldsymbol{x}_{n},\boldsymbol{p}_{n}\},n=1,2\right)=f_{1}(\boldsymbol{x}_{1},\boldsymbol{p}_{1})f_{2}(\boldsymbol{x}_{2},\boldsymbol{p}_{2})\left\langle r_{1}r_{2}\right|\hat{\rho}_{(12)}\left|s_{1}s_{2}\right\rangle ,\label{eq:wigner-12}
\end{equation}
where $f_{n}(\boldsymbol{x}_{n},\boldsymbol{p}_{n})$ with $n=1,2$
are one-particle distribution functions, and the SDM for two particles
$\hat{\rho}_{(12)}$ can be parameterized as 
\begin{equation}
\hat{\rho}_{(12)}=\hat{\rho}_{(1)}(\boldsymbol{x}_{1},\boldsymbol{p}_{1})\otimes\hat{\rho}_{(2)}(\boldsymbol{x}_{2},\boldsymbol{p}_{2})+\frac{1}{2^{2}}c_{ij}^{(12)}(\boldsymbol{x}_{1},\boldsymbol{p}_{1},\boldsymbol{x}_{2},\boldsymbol{p}_{2})\sigma_{i}^{(1)}\otimes\sigma_{j}^{(2)},\label{eq:rho-spin-12}
\end{equation}
with $i,j=x,y,z$ being the indices of Pauli matrices. Note that $\hat{\rho}_{(n)}(\boldsymbol{x}_{n},\boldsymbol{p}_{n})$
with $n=1,2$ are one-particle SDM defined in (\ref{eq:one-particle-rho})
and $c_{ij}^{(12)}(\boldsymbol{x}_{1},\boldsymbol{p}_{1},\boldsymbol{x}_{2},\boldsymbol{p}_{2})$
is the spin correlation function between two particles.

Similar to Eq. (\ref{eq:rho-12}), we can define the density operator
$\rho_{(123)}^{\mathrm{op}}$ for three particles with the joint Wigner
function similar to (\ref{eq:wigner-12}), 
\begin{equation}
W\left(\{r_{n},s_{n};\boldsymbol{x}_{n},\boldsymbol{p}_{n}\},n=1,2,3\right)=\prod_{n=1,2,3}f_{n}(\boldsymbol{x}_{n},\boldsymbol{p}_{n})\left\langle r_{1}r_{2}r_{3}\right|\hat{\rho}_{(123)}\left|s_{1}s_{2}s_{3}\right\rangle ,
\end{equation}
where the SDM for three particles is given as 
\begin{align}
\hat{\rho}_{(123)}= & \hat{\rho}^{(1)}(\boldsymbol{x}_{1},\boldsymbol{p}_{1})\otimes\hat{\rho}^{(2)}(\boldsymbol{x}_{2},\boldsymbol{p}_{2})\otimes\hat{\rho}^{(3)}(\boldsymbol{x}_{3},\boldsymbol{p}_{3})\nonumber \\
 & +\frac{1}{2^{3}}c_{ijk}^{(123)}(\boldsymbol{x}_{1},\boldsymbol{p}_{1};\boldsymbol{x}_{2},\boldsymbol{p}_{2};\boldsymbol{x}_{3},\boldsymbol{p}_{3})\sigma_{i}^{(1)}\otimes\sigma_{j}^{(2)}\otimes\sigma_{k}^{(3)}\nonumber \\
 & +\frac{1}{2^{2}}\left[c_{ij}^{(12)}(\boldsymbol{x}_{1},\boldsymbol{p}_{1};\boldsymbol{x}_{2},\boldsymbol{p}_{2})\sigma_{i}^{(1)}\otimes\sigma_{j}^{(2)}\otimes\hat{\rho}^{(3)}(\boldsymbol{x}_{3},\boldsymbol{p}_{3})\right.\nonumber \\
 & +c_{jk}^{(23)}(\boldsymbol{x}_{2},\boldsymbol{p}_{2};\boldsymbol{x}_{3},\boldsymbol{p}_{3})\hat{\rho}^{(1)}(\boldsymbol{x}_{1},\boldsymbol{p}_{1})\otimes\sigma_{j}^{(2)}\otimes\sigma_{k}^{(3)}\nonumber \\
 & \left.+c_{ik}^{(13)}(\boldsymbol{x}_{1},\boldsymbol{p}_{1};\boldsymbol{x}_{3},\boldsymbol{p}_{3})\sigma_{i}^{(1)}\otimes\hat{\rho}^{(2)}(\boldsymbol{x}_{2},\boldsymbol{p}_{2})\otimes\sigma_{k}^{(3)}\right].\label{eq:rho-spin-123}
\end{align}
Here $i,j,k=x,y,z$ are the indices of Pauli matrices, and $c_{ijk}^{(123)}$
is the spin correlation function for three particles.


\subsection{Spin Wigner function for vector mesons and baryons}

Now we can apply the density operators $\rho_{(12)}^{\mathrm{op}}$
and $\rho_{(123)}^{\mathrm{op}}$ to the quark-antiquark and three-quark
system respectively, in which we used the shorthand notations $(12)\equiv(q_{1}\bar{q}_{2})$
and $(123)\equiv(q_{1}q_{2}q_{3})$. We consider hadronization processes
$q_{1}\bar{q}_{2}\to V$ and $q_{1}q_{2}q_{3}\to B$. The spin density
operators of the vector meson and baryon are given by 
\begin{align}
\rho_{V}^{\mathrm{op}}= & \mathcal{M}_{V}\rho_{(q_{1}\bar{q}_{2})}^{\mathrm{op}}\mathcal{M}_{V}^{\dagger},\nonumber \\
\rho_{B}^{\mathrm{op}}= & \mathcal{M}_{B}\rho_{(q_{1}q_{2}q_{3})}^{\mathrm{op}}\mathcal{M}_{B}^{\dagger},\label{eq:rho-VH}
\end{align}
where $\mathcal{M}_{V/B}$ denote the transition operators for $q_{1}\bar{q}_{2}$
and $q_{1}q_{2}q_{3}$ to form the vector meson $V$ and baryon $B$
respectively. The SWF for the vector meson and baryon in phase space
can be obtained by 
\begin{equation}
\rho_{m^{\prime}m}^{V/B}(\boldsymbol{x},\boldsymbol{p})=\int[d^{3}\boldsymbol{q}]e^{i\boldsymbol{q}\cdot\boldsymbol{x}}\left\langle jm^{\prime};\boldsymbol{p}+\frac{\boldsymbol{q}}{2}\right|\rho_{V/B}^{\mathrm{op}}\left|jm;\boldsymbol{p}-\frac{\boldsymbol{q}}{2}\right\rangle ,\label{eq:rho-vb}
\end{equation}
where $j$ and $m(m^{\prime})$ denote the spin and magnetic quantum
numbers for the vector meson and baryon.

Evaluation of $\rho_{m^{\prime}m}^{V/B}$ by Eq. (\ref{eq:rho-vb})
involves the amplitudes of $\mathcal{M}$ which map the quarks to
hadrons in the forms
\begin{align}
 & \left\langle s_{1}s_{2};\boldsymbol{p}_{1}-\frac{\boldsymbol{q}_{1}}{2},\boldsymbol{p}_{2}-\frac{\boldsymbol{q}_{2}}{2}\right|\mathcal{M}_{V}^{\dagger}\left|jm;\boldsymbol{p}-\frac{\boldsymbol{q}}{2}\right\rangle \nonumber \\
\approx & \left\langle s_{1}s_{2}\right|\mathcal{M}_{V}^{\dagger\mathrm{spin}}\left|jm\right\rangle \left\langle \boldsymbol{p}_{1}-\frac{\boldsymbol{q}_{1}}{2},\boldsymbol{p}_{2}-\frac{\boldsymbol{q}_{2}}{2}\right|\mathcal{M}_{V}^{\dagger\mathrm{mom}}\left|\boldsymbol{p}-\frac{\boldsymbol{q}}{2}\right\rangle ,\label{eq:amp-v}
\end{align}
\begin{align}
 & \left\langle s_{1}s_{2}s_{3};\boldsymbol{p}_{1}-\frac{\boldsymbol{q}_{1}}{2},\boldsymbol{p}_{2}-\frac{\boldsymbol{q}_{2}}{2},\boldsymbol{p}_{3}-\frac{\boldsymbol{q}_{3}}{2}\right|\mathcal{M}_{B}^{\dagger}\left|jm;\boldsymbol{p}-\frac{\boldsymbol{q}}{2}\right\rangle \nonumber \\
\approx & \left\langle s_{1}s_{2}s_{3}\right|\mathcal{M}_{B}^{\dagger\mathrm{spin}}\left|jm\right\rangle \left\langle \boldsymbol{p}_{1}-\frac{\boldsymbol{q}_{1}}{2},\boldsymbol{p}_{2}-\frac{\boldsymbol{q}_{2}}{2},\boldsymbol{p}_{3}-\frac{\boldsymbol{q}_{3}}{2}\right|\mathcal{M}_{B}^{\dagger\mathrm{mom}}\left|\boldsymbol{p}-\frac{\boldsymbol{q}}{2}\right\rangle .\label{eq:amp-b}
\end{align}
Here we have used the factorization ansatz for the transition operators
$\mathcal{M}_{V/B}=\mathcal{M}_{V/B}^{\mathrm{spin}}\otimes\mathcal{M}_{V/B}^{\mathrm{mom}}$,
where $\mathcal{M}_{V/B}^{\mathrm{spin}}$ and $\mathcal{M}_{V/B}^{\mathrm{mom}}$
are the spin and momentum parts of $\mathcal{M}_{V/B}$ respectively.
The spin parts of the amplitudes are assumed to be proportional to
the Clebsch-Gordan coefficients 
\begin{align}
\left\langle s_{1}s_{2}\right|\mathcal{M}_{V}^{\dagger\mathrm{spin}}\left|jm\right\rangle = & N_{V}\left\langle s_{1}s_{2}\right|\left.jm\right\rangle =N_{V}C(jm;s_{1}s_{2}),\nonumber \\
\left\langle s_{1}s_{2}s_{3}\right|\mathcal{M}_{B}^{\dagger\mathrm{spin}}\left|jm\right\rangle = & N_{B}\left\langle s_{1}s_{2}s_{3}\right|\left.jm\right\rangle =N_{B}C(jm;s_{1}s_{2}s_{3}).\label{eq:cg-coeff}
\end{align}
Here $N_{V}$ and $N_{B}$ are two proportionality constants. The
momentum parts of the amplitudes are assumed to be proportional to
the vector meson's and baryon's non-relativistic wave functions 
\begin{align}
\left\langle \boldsymbol{p}_{1},\boldsymbol{p}_{2}\right|\mathcal{M}_{V}^{\dagger\mathrm{mom}}\left|\boldsymbol{p}\right\rangle = & (2\pi)^{3}\psi_{V}(\boldsymbol{p}_{b})\delta^{(3)}\left(\boldsymbol{p}-\boldsymbol{p}_{1}-\boldsymbol{p}_{2}\right),\nonumber \\
\left\langle \boldsymbol{p}_{1},\boldsymbol{p}_{2},\boldsymbol{p}_{3}\right|\mathcal{M}_{B}^{\dagger\mathrm{mom}}\left|\boldsymbol{p}\right\rangle = & (2\pi)^{3}\psi_{B}(\boldsymbol{p}_{b},\boldsymbol{p}_{c})\delta^{(3)}\left(\boldsymbol{p}-\boldsymbol{p}_{1}-\boldsymbol{p}_{2}-\boldsymbol{p}_{3}\right),\label{eq:wave}
\end{align}
where $\boldsymbol{p}_{b}\equiv(\boldsymbol{p}_{1}-\boldsymbol{p}_{2})/2$,
$\boldsymbol{p}_{c}\equiv(\boldsymbol{p}_{1}+\boldsymbol{p}_{2}-2\boldsymbol{p}_{3})/3$,
and $\psi_{V,B}$ are the vector meson's and baryon's wave functions
in momentum space.

Using Eqs. (\ref{eq:rho-spin-12}) and (\ref{eq:amp-v}) in Eq. (\ref{eq:rho-vb}),
we finally obtain the SWF for vector mesons as 
\begin{align}
\rho_{m^{\prime}m}^{V}(\boldsymbol{x},\boldsymbol{p})\approx & N_{V}^{2}\int d^{3}\boldsymbol{x}_{b}[d^{3}\boldsymbol{p}_{b}][d^{3}\boldsymbol{q}_{b}]\exp\left(-i\boldsymbol{q}_{b}\cdot\boldsymbol{x}_{b}\right)\nonumber \\
 & \times f_{q}\left(\boldsymbol{x}+\frac{1}{2}\boldsymbol{x}_{b},\frac{1}{2}\boldsymbol{p}+\boldsymbol{p}_{b}\right)f_{\bar{q}}\left(\boldsymbol{x}-\frac{1}{2}\boldsymbol{x}_{b},\frac{1}{2}\boldsymbol{p}-\boldsymbol{p}_{b}\right)\nonumber \\
 & \times\psi_{V}^{*}\left(\boldsymbol{p}_{b}+\frac{1}{2}\boldsymbol{q}_{b}\right)\psi_{V}\left(\boldsymbol{p}_{b}-\frac{1}{2}\boldsymbol{q}_{b}\right)\nonumber \\
 & \times\sum_{r_{1},s_{1},r_{2},s_{2}}C(jm;s_{1}s_{2})C^{*}(jm^{\prime};r_{1}r_{2})\left\langle r_{1}r_{2}\right|\hat{\rho}_{(q_{1}\bar{q}_{2})}\left|s_{1}s_{2}\right\rangle .\label{eq:rho-v-mm1}
\end{align}
In the elements of SDM $\left\langle r_{1}r_{2}\right|\hat{\rho}_{(q_{1}\bar{q}_{2})}\left|s_{1}s_{2}\right\rangle $,
we have set $\boldsymbol{x}_{1,2}=\boldsymbol{x}\pm\boldsymbol{x}_{b}/2$
and $\boldsymbol{p}_{1,2}=\boldsymbol{p}/2\pm\boldsymbol{p}_{b}$,
where $\boldsymbol{x}_{1,2}$ and $\boldsymbol{p}_{1,2}$ are the
positions and momenta of the quark and antiquark defined in Eq. (\ref{eq:rho-12}).
Note that the meson's wave function in momentum space is used in (\ref{eq:rho-v-mm1}).
We can rewrite (\ref{eq:rho-v-mm1}) in terms of the meson's wave
function in coordinate space as 
\begin{align}
\rho_{m^{\prime}m}^{V}(\boldsymbol{x},\boldsymbol{p})\approx & N_{V}^{2}\int d^{3}\boldsymbol{y}_{1}d^{3}\boldsymbol{y}_{2}[d^{3}\boldsymbol{p}_{b}]\exp\left[i\boldsymbol{p}_{b}\cdot(\boldsymbol{y}_{1}-\boldsymbol{y}_{2})\right]\psi_{V}^{*}\left(\boldsymbol{y}_{1}\right)\psi_{V}\left(\boldsymbol{y}_{2}\right)\nonumber \\
 & \times f_{q}\left(\boldsymbol{x}+\frac{1}{4}(\boldsymbol{y}_{1}+\boldsymbol{y}_{2}),\frac{1}{2}\boldsymbol{p}+\boldsymbol{p}_{b}\right)f_{\bar{q}}\left(\boldsymbol{x}-\frac{1}{4}(\boldsymbol{y}_{1}+\boldsymbol{y}_{2}),\frac{1}{2}\boldsymbol{p}-\boldsymbol{p}_{b}\right)\nonumber \\
 & \times\sum_{r_{1},s_{1},r_{2},s_{2}}C(jm;s_{1}s_{2})C^{*}(jm^{\prime};r_{1}r_{2})\left\langle r_{1}r_{2}\right|\hat{\rho}_{(q_{1}\bar{q}_{2})}\left|s_{1}s_{2}\right\rangle ,\label{eq:rho-v-mm2}
\end{align}
where we have used the Fourier transformation for the meson's wave
functions (the upper/lower sign corresponds to $\boldsymbol{y}_{1}$/$\boldsymbol{y}_{2}$)
\begin{equation}
\psi_{V}\left(\boldsymbol{p}_{b}\pm\frac{1}{2}\boldsymbol{q}_{b}\right)=\int d^{3}\boldsymbol{y}_{1,2}\psi_{V}(\boldsymbol{y}_{1,2})\exp\left[-i\left(\boldsymbol{p}_{b}\pm\frac{1}{2}\boldsymbol{q}_{b}\right)\cdot\boldsymbol{y}_{1,2}\right],
\end{equation}
and set $\boldsymbol{x}_{1,2}=\boldsymbol{x}\pm(\boldsymbol{y}_{1}+\boldsymbol{y}_{2})/4$
and $\boldsymbol{p}_{1,2}=\boldsymbol{p}/2\pm\boldsymbol{p}_{b}$
in all phase space functions in $\left\langle r_{1}r_{2}\right|\hat{\rho}_{(q_{1}\bar{q}_{2})}\left|s_{1}s_{2}\right\rangle $.

Similarly, using Eqs. (\ref{eq:rho-spin-123}) and (\ref{eq:amp-b})
in Eq. (\ref{eq:rho-vb}), we can derive the SWF for baryons as 
\begin{align}
\rho_{m^{\prime}m}^{B}(\boldsymbol{x},\boldsymbol{p})\approx & N_{B}^{2}\int\prod_{n=b,c}[d^{3}\boldsymbol{p}_{n}][d^{3}\boldsymbol{q}_{n}]d^{3}\boldsymbol{x}_{n}\nonumber \\
 & \times\exp\left(-i\boldsymbol{q}_{b}\cdot\boldsymbol{x}_{b}-i\boldsymbol{q}_{c}\cdot\boldsymbol{x}_{c}\right)f_{q_{1}}(\boldsymbol{x}_{1},\boldsymbol{p}_{1})f_{q_{2}}(\boldsymbol{x}_{2},\boldsymbol{p}_{2})f_{q_{3}}(\boldsymbol{x}_{3},\boldsymbol{p}_{3})\nonumber \\
 & \times\psi_{B}^{*}\left(\boldsymbol{p}_{b}+\frac{1}{2}\boldsymbol{q}_{b},\boldsymbol{p}_{c}+\frac{1}{2}\boldsymbol{q}_{c}\right)\psi_{B}\left(\boldsymbol{p}_{b}-\frac{1}{2}\boldsymbol{q}_{b},\boldsymbol{p}_{c}-\frac{1}{2}\boldsymbol{q}_{c}\right)\nonumber \\
 & \times\sum_{\{r_{n},s_{n},n=1,2,3\}}C(jm;s_{1}s_{2}s_{3})C^{*}(jm^{\prime};r_{1}r_{2}r_{3})\nonumber \\
 & \times\left\langle r_{1}r_{2}r_{3}\right|\hat{\rho}_{(q_{1}q_{2}q_{3})}\left|s_{1}s_{2}s_{3}\right\rangle ,\label{eq:rho-baryon}
\end{align}
where we have set the positions and momenta of three quarks in the
distribution functions and in $\left\langle r_{1}r_{2}r_{3}\right|\hat{\rho}_{(q_{1}q_{2}q_{3})}\left|s_{1}s_{2}s_{3}\right\rangle $
as 
\begin{align}
\boldsymbol{x}_{1}= & \boldsymbol{x}+\frac{1}{2}\boldsymbol{x}_{b}+\frac{1}{3}\boldsymbol{x}_{c},\nonumber \\
\boldsymbol{x}_{2}= & \boldsymbol{x}-\frac{1}{2}\boldsymbol{x}_{b}+\frac{1}{3}\boldsymbol{x}_{c},\nonumber \\
\boldsymbol{x}_{3}= & \boldsymbol{x}-\frac{2}{3}\boldsymbol{x}_{c},\label{eq:3q-position}\\
\boldsymbol{p}_{1}= & \frac{1}{3}\boldsymbol{p}+\boldsymbol{p}_{b}+\frac{1}{2}\boldsymbol{p}_{c},\nonumber \\
\boldsymbol{p}_{2}= & \frac{1}{3}\boldsymbol{p}-\boldsymbol{p}_{b}+\frac{1}{2}\boldsymbol{p}_{c},\nonumber \\
\boldsymbol{p}_{3}= & \frac{1}{3}\boldsymbol{p}-\boldsymbol{p}_{c}.\label{eq:3q-mom}
\end{align}
We can also rewrite (\ref{eq:rho-baryon}) in terms of the baryon's
wave functions in coordiante space as 
\begin{align}
\rho_{m^{\prime}m}^{B}(\boldsymbol{x},\boldsymbol{p})\approx & N_{B}^{2}\int d^{3}\boldsymbol{y}_{1}d^{3}\boldsymbol{y}_{2}d^{3}\boldsymbol{z}_{1}d^{3}\boldsymbol{z}_{2}[d^{3}\boldsymbol{p}_{b}][d^{3}\boldsymbol{p}_{c}]\nonumber \\
 & \times\exp\left[i\boldsymbol{p}_{b}\cdot(\boldsymbol{y}_{1}-\boldsymbol{y}_{2})+i\boldsymbol{p}_{c}\cdot(\boldsymbol{z}_{1}-\boldsymbol{z}_{2})\right]\nonumber \\
 & \times f_{q_{1}}(\boldsymbol{x}_{1},\boldsymbol{p}_{1})f_{q_{2}}(\boldsymbol{x}_{2},\boldsymbol{p}_{2})f_{q_{3}}(\boldsymbol{x}_{3},\boldsymbol{p}_{3})\psi_{B}^{*}\left(\boldsymbol{y}_{1},\boldsymbol{z}_{1}\right)\psi_{B}\left(\boldsymbol{y}_{2},\boldsymbol{z}_{2}\right)\nonumber \\
 & \times\sum_{\{r_{n},s_{n},n=1,2,3\}}C(jm;s_{1}s_{2}s_{3})C^{*}(jm^{\prime};r_{1}r_{2}r_{3})\left\langle r_{1}r_{2}r_{3}\right|\hat{\rho}_{(q_{1}q_{2}q_{3})}\left|s_{1}s_{2}s_{3}\right\rangle ,\label{eq:rho-baryon-1}
\end{align}
where we have set the positions of three quarks in the distribution
functions and in $\left\langle r_{1}r_{2}r_{3}\right|\hat{\rho}_{(q_{1}q_{2}q_{3})}\left|s_{1}s_{2}s_{3}\right\rangle $
as
\begin{align}
\boldsymbol{x}_{1}= & \boldsymbol{x}+\frac{1}{4}(\boldsymbol{y}_{1}+\boldsymbol{y}_{2})+\frac{1}{6}(\boldsymbol{z}_{1}+\boldsymbol{z}_{2}),\nonumber \\
\boldsymbol{x}_{2}= & \boldsymbol{x}-\frac{1}{4}(\boldsymbol{y}_{1}+\boldsymbol{y}_{2})+\frac{1}{6}(\boldsymbol{z}_{1}+\boldsymbol{z}_{2}),\nonumber \\
\boldsymbol{x}_{3}= & \boldsymbol{x}-\frac{1}{3}(\boldsymbol{z}_{1}+\boldsymbol{z}_{2}),\label{eq:position-3q}
\end{align}
while the momenta of three quarks are the same as in Eq. (\ref{eq:3q-mom}).
We also used the Fourier transformation to convert the baryon's wave
functions from momentum space to coordinate space 
\begin{align}
 & \psi_{B}\left(\boldsymbol{p}_{b}\pm\frac{1}{2}\boldsymbol{q}_{b},\boldsymbol{p}_{c}\pm\frac{1}{2}\boldsymbol{q}_{c}\right)\nonumber \\
= & \int d^{3}\boldsymbol{y}_{1,2}d^{3}\boldsymbol{z}_{1,2}\psi_{B}\left(\boldsymbol{y}_{1,2},\boldsymbol{z}_{1,2}\right)\nonumber \\
 & \exp\left[-i\left(\boldsymbol{p}_{b}\pm\frac{1}{2}\boldsymbol{q}_{b}\right)\cdot\boldsymbol{y}_{1,2}-i\left(\boldsymbol{p}_{c}\pm\frac{1}{2}\boldsymbol{q}_{c}\right)\cdot\boldsymbol{z}_{1,2}\right],
\end{align}
where the upper/lower sign corresponds to $(\boldsymbol{y}_{1},\boldsymbol{z}_{1})$/$(\boldsymbol{y}_{2},\boldsymbol{z}_{2})$.


\subsection{Short- and long-range correlations \label{subsec:short-long-corr}}

We have finally derived in Eqs. (\ref{eq:rho-v-mm2}) and (\ref{eq:rho-baryon-1})
the SWF for the vector meson and baryon in phase space, which are
the main results of this section. In the SWF for the vector meson
given by (\ref{eq:rho-v-mm2}), we see integrals over coordinates
$\boldsymbol{y}_{1}$ and $\boldsymbol{y}_{2}$ in the vector meson's
wave functions $\psi_{V}^{*}(\boldsymbol{y}_{1})\psi_{V}(\boldsymbol{y}_{2})$
and also an integral over $\boldsymbol{p}_{b}$ (the relative momentum
of $q$ and $\bar{q}$) which is conjugate to $\boldsymbol{y}_{1}-\boldsymbol{y}_{2}$.
In the integrand, there are phase space distributions for $q$ and
$\bar{q}$ which depend on $\boldsymbol{y}_{1}+\boldsymbol{y}_{2}$
and $\boldsymbol{p}_{b}$. As part of the integrand, the polarization
functions $\boldsymbol{P}_{q}(\boldsymbol{x}_{1},\boldsymbol{p}_{1})$,
$\boldsymbol{P}_{\bar{q}}(\boldsymbol{x}_{2},\boldsymbol{p}_{2})$
and correlation function $c_{ij}^{(12)}(\boldsymbol{x}_{1},\boldsymbol{p}_{1},\boldsymbol{x}_{2},\boldsymbol{p}_{2})$
or any product of them appear in the SDM. In other words, the SWF
for the vector meson involves an average of the polarization and correlation
functions for $q$ and $\bar{q}$ over the phase space volume defined
by the meson's wave function and weighted by the distribution functions
of $q$ and $\bar{q}$. So the averages (or integrals) over $\boldsymbol{y}_{1}$,
$\boldsymbol{y}_{2}$ and $\boldsymbol{p}_{b}$ is restricted by the
meson's wave function and correspond to the short-range correlation
for $q$ and $\bar{q}$. In contrast, an average over the phase space
variables $(\boldsymbol{x},\boldsymbol{p})$ of the vector meson corresponds
to the long-range correlation outside the meson's internal space. 


Similar to the case of the vector meson, there are also short- and
long-range correlations of three quarks in the baryon. The only difference
from the meson's case is that the short-range correlation in the baryon
involves the integrals over $(\boldsymbol{y}_{1},\boldsymbol{z}_{1})$,
$(\boldsymbol{y}_{2},\boldsymbol{z}_{2})$, $\boldsymbol{p}_{b}$
(conjugate to $\boldsymbol{y}_{1}-\boldsymbol{y}_{2}$) and $\boldsymbol{p}_{c}$
(conjugate to $\boldsymbol{z}_{1}-\boldsymbol{z}_{2}$) in the baryon's
SWF. This is because the baryon's wave function depends on two relative
coordinates $\boldsymbol{y}$ and $\boldsymbol{z}$ of three quarks
as shown in (\ref{eq:position-3q}). 


We can rewrite Eqs. (\ref{eq:rho-v-mm2}) and (\ref{eq:rho-baryon-1})
into compact forms: 
\begin{align}
\rho_{m^{\prime}m}^{V}(\boldsymbol{x},\boldsymbol{p})\approx & N_{V}^{2}\sum_{r_{1},s_{1},r_{2},s_{2}}C(jm;s_{1}s_{2})C^{*}(jm^{\prime};r_{1}r_{2})\nonumber \\
 & \times\left\langle r_{1}r_{2}\right|\left\langle \hat{\rho}_{(q_{1}\bar{q}_{2})}(\boldsymbol{x}_{1},\boldsymbol{p}_{1},\boldsymbol{x}_{2},\boldsymbol{p}_{2})\right\rangle _{V}\left|s_{1}s_{2}\right\rangle ,\label{eq:rho-v-sra}\\
\rho_{m^{\prime}m}^{B}(\boldsymbol{x},\boldsymbol{p})\approx & N_{B}^{2}\sum_{\{r_{n},s_{n},n=1,2,3\}}C(jm;s_{1}s_{2}s_{3})C^{*}(jm^{\prime};r_{1}r_{2}r_{3})\nonumber \\
 & \times\left\langle r_{1}r_{2}r_{3}\right|\left\langle \hat{\rho}_{(q_{1}q_{2}q_{3})}(\boldsymbol{x}_{1},\boldsymbol{p}_{1},\boldsymbol{x}_{2},\boldsymbol{p}_{2},\boldsymbol{x}_{3},\boldsymbol{p}_{3})\right\rangle _{B}\left|s_{1}s_{2}s_{3}\right\rangle ,\label{eq:rho-b-sra}
\end{align}
where $\left\langle \hat{\rho}_{(q_{1}\bar{q}_{2})}\right\rangle _{V}$
and $\left\langle \hat{\rho}_{(q_{1}q_{2}q_{3})}\right\rangle _{B}$
represent the averages (integrals) in Eqs. (\ref{eq:rho-v-mm2}) and
(\ref{eq:rho-baryon-1}) over phase space volumes inside the meson
and baryon respectively that give the short-range correlations.


\subsection{Hadronization process as a quantum measurement}

In Quantum Information Science (QIS), a quantum measurement process
is in general made through a collection of measurement operators satisfying
a completeness relation, which define a generalized measurement \citep{Nielsen2015}.
\textit{\emph{Projective (von Neumann) measurements arise as a special
case in which the measurement operators are orthogonal projectors.
In Ref. \citep{Wu:2024mtj}, the decays of hyperons are described
as effective generalized measurements acting on the spin degrees of
freedom. In the following we argue that the hadronization process
can also be interpreted as a type of generalized measurement.}}

The key observation is that the transition operators $\mathcal{M}_{V/B}=\mathcal{M}_{V/B}^{\mathrm{spin}}\otimes\mathcal{M}_{V/B}^{\mathrm{mom}}$
in (\ref{eq:rho-VH}) play the role of measurement operators. The
effects of $\mathcal{M}_{V/B}$ are shown in Eqs. (\ref{eq:rho-v-mm2})
and (\ref{eq:rho-baryon-1}) or in compact forms in Eqs. (\ref{eq:rho-v-sra})
and (\ref{eq:rho-b-sra}). We can look at a very simple case in which
the space and momentum dependence of all variables is neglected. Then
the spin density matrices for vector mesons and baryons in Eqs. (\ref{eq:rho-v-sra})
and (\ref{eq:rho-b-sra}) can be put into simple forms (we neglected
the normalization of spin density matrices which will be done later):
\begin{align}
\rho_{m^{\prime}m}^{V}= & \sum_{r_{1},s_{1},r_{2},s_{2}}C(jm;s_{1}s_{2})C^{*}(jm^{\prime};r_{1}r_{2})\left\langle r_{1}r_{2}\right|\hat{\rho}_{(q_{1}\bar{q}_{2})}\left|s_{1}s_{2}\right\rangle ,\nonumber \\
\rho_{m^{\prime}m}^{B}= & \sum_{\{r_{n},s_{n},n=1,2,3\}}C(jm;s_{1}s_{2}s_{3})C^{*}(jm^{\prime};r_{1}r_{2}r_{3})\nonumber \\
 & \times\left\langle r_{1}r_{2}r_{3}\right|\hat{\rho}_{(q_{1}q_{2}q_{3})}\left|s_{1}s_{2}s_{3}\right\rangle ,\label{eq:rho-v-b}
\end{align}
where $j=1$ and $m=\pm1,0$ in $\rho_{m^{\prime}m}^{V}$ for the
vector meson, and $j=1/2$ or $3/2$ and $m=\pm1/2$ or $m=\pm1/2,\pm3/2$
in $\rho_{m^{\prime}m}^{B}$ for the baryon. We see that the Clebsch-Gordan
coefficients are actually measurement operators (matrices) in spin
space 
\begin{align}
\mathcal{M}_{M}^{(j)}(m;r_{1}r_{2})\equiv & C^{*}(jm;r_{1}r_{2}),\nonumber \\
\mathcal{M}_{B}^{(j)}(m;r_{1}r_{2}r_{3})\equiv & C^{*}(jm;r_{1}r_{2}r_{3}),\label{eq:measure-op}
\end{align}
where the subscript ``$M$'' in $\mathcal{M}_{M}^{(j)}$ stands
for mesons with $j=0,1$ correpsonding to the scalar and vector meson
respectively. Note that the measurement operator for the vector meson
is denoted as $\mathcal{M}_{V}^{(j)}\equiv\mathcal{M}_{M}^{(j=1)}$.
So $\mathcal{M}_{M}^{(j)}$ is a $(2j+1)\times4$ matrix and $\mathcal{M}_{B}^{(j)}$
is a $(2j+1)\times8$ matrix, while $\mathcal{M}_{M}^{(j)\dagger}$
is a $4\times(2j+1)$ matrix and $\mathcal{M}_{B}^{(j)\dagger}$ is
a $8\times(2j+1)$ matrix. 

With the measurement operators (\ref{eq:measure-op}) and their Hermitian
conjugates, Eq. (\ref{eq:rho-v-b}) can be put into the compact form
\begin{align}
\rho_{V}= & \mathcal{M}_{V}^{(j)}\hat{\rho}_{(q_{1}\bar{q}_{2})}\mathcal{M}_{V}^{(j)\dagger},\nonumber \\
\rho_{B}= & \mathcal{M}_{B}^{(j)}\hat{\rho}_{(q_{1}q_{2}q_{3})}\mathcal{M}_{B}^{(j)\dagger},
\end{align}
where we have put the superscripts ``$V$'' and ``$B$'' in Eq.
(\ref{eq:rho-v-b}) to subscripts for notational convenience. The
probabilities of obtaining the vector meson and the spin-$j$ baryon
are 
\begin{align}
\mathcal{P}_{M}(j=1)= & \textrm{Tr}\rho_{V}=\textrm{Tr}\left[\mathcal{M}_{V}^{(j)}\hat{\rho}_{(q_{1}\bar{q}_{2})}\mathcal{M}_{V}^{(j)\dagger}\right],\nonumber \\
\mathcal{P}_{B}(j)= & \textrm{Tr}\rho_{B}=\textrm{Tr}\left[\mathcal{M}_{B}^{(j)}\hat{\rho}_{(q_{1}q_{2}q_{3})}\mathcal{M}_{B}^{(j)\dagger}\right],
\end{align}
where the traces are taken in spin space. Note that $\left\{ \mathcal{P}_{M}(j)\right\} $
and $\left\{ \mathcal{P}_{B}(j)\right\} $ are probabilities for all
possible $j$ for mesons and baryons with the normalization conditions
\begin{align}
\sum_{j=0,1}\mathcal{P}_{M}(j)= & 1,\nonumber \\
\sum_{j=1/2,1/2,3/2}\mathcal{P}_{B}(j)= & 1.
\end{align}
For the measurement operators, we have the properties of the positive
semidefiniteness 
\begin{equation}
\mathcal{M}_{M/B}^{(j)\dagger}\mathcal{M}_{M/B}^{(j)}\geq0,
\end{equation}
and the completeness 
\begin{align}
\sum_{j=0,1}\mathcal{M}_{M}^{(j)\dagger}\mathcal{M}_{M}^{(j)}= & 1_{4},\nonumber \\
\sum_{j=1/2,1/2,3/2}\mathcal{M}_{B}^{(j)\dagger}\mathcal{M}_{B}^{(j)}= & 1_{8},
\end{align}
where $1_{n}$ denotes the $n\times n$ unit matrix, and there are
two spin doublets (corresponding to the spin triplet and singlet of
two quarks) in forming the baryon from three quarks. Based on the
measurement postulate in quantum information science, by the generalized
measurement, the initial states $\hat{\rho}_{(q_{1}\bar{q}_{2})}$
and $\hat{\rho}_{(q_{1}q_{2}q_{3})}$ instantaneously transform to
the states $\hat{\rho}_{M}^{(j)}$ and $\hat{\rho}_{B}^{(j)}$ respectively
\begin{align}
\hat{\rho}_{(q_{1}\bar{q}_{2})}\mapsto\hat{\rho}_{M}^{(j)}\equiv & \frac{1}{\mathcal{P}_{M}(j)}\mathcal{M}_{M}^{(j)}\hat{\rho}_{(q_{1}\bar{q}_{2})}\mathcal{M}_{M}^{(j)\dagger},\nonumber \\
\hat{\rho}_{(q_{1}q_{2}q_{3})}\mapsto\hat{\rho}_{B}^{(j)}\equiv & \frac{1}{\mathcal{P}_{B}(j)}\mathcal{M}_{B}^{(j)}\hat{\rho}_{(q_{1}q_{2}q_{3})}\mathcal{M}_{B}^{(j)\dagger}.
\end{align}
Note that $\hat{\rho}_{M}^{(j)}$ and $\hat{\rho}_{B}^{(j)}$ are
normalized so that we have $\mathrm{Tr}\hat{\rho}_{M}^{(j)}=1$ and
$\mathrm{Tr}\hat{\rho}_{B}^{(j)}=1$.

We can calculate the von Neumann entropy for $q_{1}\bar{q}_{2}$ and
$q_{1}q_{2}q_{3}$ as well as for the meson and baryon through their
normalized SDMs as 
\begin{equation}
S\left(\hat{\rho}_{i}\right)=-\mathrm{Tr}\left(\hat{\rho}_{i}\ln\hat{\rho}_{i}\right),
\end{equation}
where $i=q_{1}\bar{q}_{2},q_{1}q_{2}q_{3},M,B$. We can rewrite $\hat{\rho}$
($\hat{\rho}\equiv\hat{\rho}_{i}$, we suppress the index $i$) as
$\hat{\rho}=\hat{\rho}_{0}+\delta\hat{\rho}$, where $\hat{\rho}_{0}=1_{D}/D$
with $1_{D}$ being the unit matrix of dimension $D$. When the matrix
elements $\delta\hat{\rho}_{mn}\ll1$ we can approximate the von Neumann
entropy to the quadratic order $(\delta\hat{\rho})^{2}$ 
\begin{equation}
S\left(\hat{\rho}\right)\approx\ln D-\frac{D}{2}\mathrm{Tr}\left[(\delta\hat{\rho})^{2}\right].\label{eq:vn-entropy}
\end{equation}
For a maximall-mixed state we have $\mathrm{Tr}(\hat{\rho}^{2})=\mathrm{Tr}(\hat{\rho}_{0}^{2})=1/D$,
while for a pure state we have $\mathrm{Tr}(\hat{\rho}^{2})=1$. Then
we obtain the inequality 
\begin{equation}
0\leq\mathrm{Tr}\left[(\delta\hat{\rho})^{2}\right]\leq1-1/D.\label{eq:inequality-rho2}
\end{equation}
So we see that the second term of the von Neumann entropy (\ref{eq:vn-entropy})
is always negative.

Using Eq. (\ref{eq:vn-entropy}) we can calculate $S\left(\hat{\rho}_{(q_{1}\bar{q}_{2})}\right)$
and $S\left(\hat{\rho}_{V}\right)$ to the quadratic order of the
spin density's deviation from $\hat{\rho}_{0}$. By assuming $P_{q_{1}/\bar{q}_{2}}\equiv\sqrt{\boldsymbol{P}_{q_{1}/\bar{q}_{2}}\cdot\boldsymbol{P}_{q_{1}/\bar{q}_{2}}}\ll1$,
$c_{ij}^{(q_{1}\bar{q}_{2})}\ll1$ and $P_{q_{1}/\bar{q}_{2}}\sim c_{ij}^{(q_{1}\bar{q}_{2})}$
in Eq. (\ref{eq:rho-spin-12}) we obtain to the quadratic order in
$P_{q_{1}/\bar{q}_{2}}$ and $c_{ij}^{(q_{1}\bar{q}_{2})}$ 
\begin{align}
S\left(\hat{\rho}_{(q_{1}\bar{q}_{2})}\right)\approx & 2\ln2-\frac{1}{2}\left[P_{q_{1}}^{2}+P_{\bar{q}_{2}}^{2}+\mathrm{Tr}\left(c^{(q_{1}\bar{q}_{2})}c^{(q_{1}\bar{q}_{2})T}\right)\right],\nonumber \\
S\left(\hat{\rho}_{V}\right)\approx & \ln3-\frac{1}{3}\left[\left(\boldsymbol{P}_{q_{1}}+\boldsymbol{P}_{\bar{q}_{2}}\right)^{2}\right.\nonumber \\
 & \left.+\mathrm{Tr}\left(c^{(q_{1}\bar{q}_{2})}c^{(q_{1}\bar{q}_{2})}\right)+\mathrm{Tr}\left(c^{(q_{1}\bar{q}_{2})}c^{(q_{1}\bar{q}_{2})T}\right)-\frac{2}{3}\left(\mathrm{Tr}c^{(q_{1}\bar{q}_{2})}\right)^{2}\right],\label{eq:entropy-qv}
\end{align}
where the superscript ``$T$'' denotes the transpose of a matrix,
so we have $c_{ij}^{(q_{1}\bar{q}_{2})T}=c_{ji}^{(q_{1}\bar{q}_{2})}$.
By the inequality (\ref{eq:inequality-rho2}), one can show 
\begin{align}
\mathrm{Tr}\left(c^{(q_{1}\bar{q}_{2})}c^{(q_{1}\bar{q}_{2})T}\right)\geq & 0,\nonumber \\
\mathrm{Tr}\left(c^{(q_{1}\bar{q}_{2})}c^{(q_{1}\bar{q}_{2})}\right)+\mathrm{Tr}\left(c^{(q_{1}\bar{q}_{2})}c^{(q_{1}\bar{q}_{2})T}\right)-\frac{2}{3}\left(\mathrm{Tr}c^{(q_{1}\bar{q}_{2})}\right)^{2}\geq & 0.\label{eq:ineq-c}
\end{align}
So we see from Eq. (\ref{eq:entropy-qv}) that non-zero quark polarizations
or correlations always lower the entropy. It is easy to check the
inequalities in (\ref{eq:ineq-c}) by rewriting $c^{(12)}\equiv c^{(q_{1}\bar{q}_{2})}$
as a sum over symmetric (S) and anti-symmetric (A) part 
\begin{equation}
c^{(12)}=c_{S}^{(12)}+c_{A}^{(12)},
\end{equation}
where $c_{S,ij}^{(12)}=c_{S,ji}^{(12)}$ and $c_{A,ij}^{(12)}=-c_{A,ji}^{(12)}$.
Then the inequalities in (\ref{eq:ineq-c}) now become 
\begin{align}
\mathrm{Tr}\left(c_{S}^{(12)}\right)^{2}-\mathrm{Tr}\left(c_{A}^{(12)}\right)^{2}\geq & 0,\nonumber \\
F\left[c^{(12)}\right]\equiv\mathrm{Tr}\left(c_{S}^{(12)}\right)^{2}-\frac{1}{3}\left(\mathrm{Tr}c_{S}^{(12)}\right)^{2}\geq & 0,
\end{align}
which hold obviously.

In a similar way, we can also calculate with Eq. (\ref{eq:vn-entropy})
$S\left(\hat{\rho}_{(q_{1}q_{2}q_{3})}\right)$ and $S\left(\hat{\rho}_{B}\right)$
to the quadratic order of the spin density's deviation from $\hat{\rho}_{0}$.
The results are 
\begin{align}
S\left(\hat{\rho}_{(q_{1}q_{2}q_{3})}\right)\approx & 3\ln2-\frac{1}{2}\left[P_{q_{1}}^{2}+P_{q_{2}}^{2}+P_{q_{3}}^{2}\right.\nonumber \\
 & \left.+\mathrm{Tr}\left(c^{(12)}c^{(12)T}\right)+\mathrm{Tr}\left(c^{(23)}c^{(23)T}\right)+\mathrm{Tr}\left(c^{(13)}c^{(13)T}\right)\right],\nonumber \\
S\left(\hat{\rho}_{(3/2)B}\right)\approx & \ln4-\frac{5}{18}\left(\boldsymbol{P}_{q_{1}}+\boldsymbol{P}_{q_{2}}+\boldsymbol{P}_{q_{3}}\right)^{2}-\frac{1}{3}F\left[c^{(12)}+c^{(23)}+c^{(13)}\right],\nonumber \\
S\left(\hat{\rho}_{(1/2)B}^{MS}\right)\approx & \ln2-\frac{1}{18}\left(2\boldsymbol{P}_{q_{1}}+2\boldsymbol{P}_{q_{2}}-\boldsymbol{P}_{q_{3}}\right)^{2},\nonumber \\
S\left(\hat{\rho}_{(1/2)B}^{MA}\right)\approx & \ln2-\frac{1}{2}\boldsymbol{P}_{q_{3}}^{2},\label{eq:baryon-entropy}
\end{align}
where ``MS'' and ``MA'' mean the mixed symmetric (spin triplet)
and anti-symmetric (spin singlet) state respectively, for $q_{1}$
and $q_{2}$. Among ground states of octet baryons with spin-partity
$(1/2)^{+}$, the proton, neutron, $\Sigma^{\pm}$ and $\Xi^{-,0}$
are in mixed symmetric states with the flavor configuration $q_{1}q_{2}q_{3}=aab$
(flavor $a\neq b$), $\Sigma^{0}$ is also in the mixed symmetric
state with the flavor configuration $q_{1}q_{2}q_{3}=uds$, while
the $\Lambda$ hyperon is in the mixed anti-symmetric state with the
flavor configuration $q_{1}q_{2}q_{3}=uds$. We observe in Eq. (\ref{eq:baryon-entropy})
that there are no spin correlation terms in the entropy for octet
baryons at the quadratic order since they are of cubic or higher order,
while there are spin correlation terms for decuplet baryons. This
means that decuplet baryons are more entangled than octet baryons
in case of no quark polarization.

The comparison of the entropies of $q_{1}\bar{q}_{2}$ and $q_{1}q_{2}q_{3}$
with those of the corresponding meson and baryon states can provide
insights into the role of polarization and spin correlations in the
entropy evolution associated with quark recombination \citep{Greco:2003xt,Fries:2003vb,Greco:2003mm,Fries:2003kq,Hwa:2003bn,Shao:2004cn}
in the hadronization process of relativistic heavy-ion collisions.
It can also shed light on the entropy puzzle in the quark combination
model \citep{Song:2010bi}. Note that the discussion in this subsection
is based on the simplified SDM for the vector meson and baryon in
which we neglected the space and momentum dependence of all variables.
By using the SWFs for the vector meson and baryon in Eqs. (\ref{eq:rho-v-sra})
and (\ref{eq:rho-b-sra}), the von Neumann entropies for the quark
and hadron systems should have more sophisticated features which will
be reserved for a future study.


\section{Spin polarizations of vector mesons and baryons}

From the SWF for the vector meson and baryon we can obtain their spin
polarizations \citep{Choi:1989yf,Kim:1992az,Becattini:2016gvu,Zhao:2022lbw},
\begin{equation}
\boldsymbol{P}=\frac{1}{S}\frac{\mathrm{Tr}\left(\rho\boldsymbol{S}\right)}{\mathrm{Tr}\rho},\label{eq:polar-tr-s}
\end{equation}
where $\boldsymbol{S}$ is the spin operator (matrix) for the vector
meson or baryon, $S$ is the spin quantum number with the eigenvalue
of $\boldsymbol{S}^{2}$ being $S(S+1)$, the polarization has an
upper bound, $P\equiv\sqrt{\boldsymbol{P}\cdot\boldsymbol{P}}\leq1$.
Here $\rho$ denotes the SWF as a spin density matrix given in Eq.
(\ref{eq:rho-v-mm2}) for the vector meson and Eq. (\ref{eq:rho-baryon-1})
for the baryon. If we choose the $z$-direction as the spin quantization
direction, the spin polarizations for the vector meson, spin-1/2 and
-3/2 baryons are given by 
\begin{align}
P_{V}^{z}= & \frac{\rho_{11}-\rho_{-1,-1}}{\rho_{11}+\rho_{-1,-1}+\rho_{00}},\nonumber \\
P_{B_{1/2}}^{z}= & \frac{\rho_{1/2,1/2}-\rho_{-1/2,-1/2}}{\rho_{1/2,1/2}+\rho_{-1/2,-1/2}},\nonumber \\
P_{B_{3/2}}^{z}= & \frac{2}{3}\frac{\sum_{s_{z}=\pm3/2,\pm1/2}S_{z}\rho_{S_{z},S_{z}}}{\sum_{s_{z}=\pm3/2,\pm1/2}\rho_{S_{z},S_{z}}}\nonumber \\
= & \frac{\rho_{3/2,3/2}-\rho_{-3/2,-3/2}+\frac{1}{3}\left(\rho_{1/2,1/2}-\rho_{-1/2,-1/2}\right)}{\rho_{3/2,3/2}+\rho_{-3/2,-3/2}+\rho_{1/2,1/2}+\rho_{-1/2,-1/2}},\label{eq:polar-v-b}
\end{align}
where we have suppressed the indices $V$, $B_{1/2}$ and $B_{3/2}$
in $\rho$ for the vector meson, spin-1/2 and spin-3/2 baryons respectively.
Note that the spin polarizations in Eq. (\ref{eq:polar-v-b}) are
functions of phase space variables of mesons and baryons before taking
long-range average. The spin polarizations of hyperons can be measured
through their weak decays, but vector mesons only decay through strong
interaction so their polarizations cannot be measured in a conventional
way.


Using the spin-flavor or SU(6) wave functions of mesons and baryons
\citep{Close:1979bt}, we obtain the spin polarizations for the spin-1/2
baryons $\Xi^{-}(ssd)$, $\Xi^{0}(ssu)$ and $\Lambda(usd)$ \citep{Lv:2024uev}
and the spin-3/2 baryon $\Omega^{-}(sss)$ \citep{Zhang:2024hyq}
including spin correlations among their constituent quarks
\begin{align}
P_{\Xi^{-}}^{z}= & \frac{1}{\left\langle C_{\Xi^{-}}\right\rangle _{B}}\left\langle \frac{1}{3}\left(4P_{s}^{z}-P_{d}^{z}\right)C_{\Xi^{-}}+\delta\rho_{\Xi^{-}}\right\rangle _{B},\nonumber \\
P_{\Xi^{0}}^{z}= & P_{\Xi^{-}}(d\rightarrow u),\nonumber \\
P_{\varLambda}^{z}= & \frac{1}{\left\langle C_{\Lambda}\right\rangle _{B}}\left\langle P_{s}^{z}C_{\Lambda}-\delta\rho_{\Lambda}\right\rangle _{B},\nonumber \\
P_{\Omega}^{z}= & \frac{1}{\left\langle C_{\Omega}\right\rangle _{B}}\left\langle \frac{5}{3}P_{s}^{z}C_{\Omega}+\delta\rho_{\Omega}\right\rangle _{B},\label{eq:av-baryons}
\end{align}
where 
\begin{align}
\delta\rho_{\Xi^{-}}= & -\frac{4}{3}\left(\boldsymbol{P}_{s}\cdot\boldsymbol{P}_{s}+c_{ii}^{(ss)}-\boldsymbol{P}_{s}\cdot\boldsymbol{P}_{d}-c_{ii}^{(sd)}\right)\left(P_{s}^{z}-P_{d}^{z}\right)\nonumber \\
 & -4c_{iz}^{(ss)}P_{d}^{i}+2\left(c_{iz}^{(sd)}-2c_{zi}^{(sd)}\right)P_{s}^{i}+c_{iiz}^{(ssd)}-4c_{zii}^{(ssd)},\nonumber \\
C_{\Xi^{-}}= & 3+\boldsymbol{P}_{s}\cdot\boldsymbol{P}_{s}+c_{ii}^{(ss)}-4\left(\boldsymbol{P}_{s}\cdot\boldsymbol{P}_{d}+c_{ii}^{(sd)}\right),\nonumber \\
\delta\rho_{\Xi^{0}}= & \delta\rho_{\Xi^{-}}(d\rightarrow u),\nonumber \\
C_{\Xi^{0}}= & C_{\Xi^{-}}(d\rightarrow u),\nonumber \\
\delta\rho_{\Lambda}= & c_{iz}^{(us)}P_{d}^{i}+c_{iz}^{(ds)}P_{u}^{i}+c_{iiz}^{(uds)},\nonumber \\
C_{\Lambda}= & 1-\left(\boldsymbol{P}_{u}\cdot\boldsymbol{P}_{d}+c_{ii}^{(ud)}\right),\nonumber \\
\delta\rho_{\Omega}= & c_{zii}^{(sss)}+2c_{zi}^{(ss)}P_{s}^{i}-4\left(\boldsymbol{P}_{s}\cdot\boldsymbol{P}_{s}+c_{ii}^{(ss)}\right)P_{s}^{z},\nonumber \\
C_{\Omega}= & 3+3\left(\boldsymbol{P}_{s}\cdot\boldsymbol{P}_{s}+c_{ii}^{(ss)}\right).
\end{align}
In Eq. (\ref{eq:av-baryons}) the short-range average is taken inside
a phase space volume defined by a specific baryon's wave function
and weighted by the distribution functions of the baryon's constituent
quarks as described in Eq. (\ref{eq:rho-b-sra}) and Sect. \ref{subsec:short-long-corr}.

For the following discussion, we assume that all baryon wave functions
in coordinate or momentum space have a similar form at the hadronic
scale, which provides a good approximation for the present problem.
In other words, we adopt the same wave-function form for all baryon
species.


The 00-element of the spin density matrix for the $\phi$ meson is
usually called spin alignment and given by 
\begin{equation}
\rho_{00}^{\phi}=\frac{1}{\left\langle C_{\phi}\right\rangle _{V}}\left[1+\left\langle \boldsymbol{P}_{s}\cdot\boldsymbol{P}_{\bar{s}}+c_{ii}^{(s\bar{s})}-2\left(P_{s}^{z}P_{\bar{s}}^{z}+c_{zz}^{(s\bar{s})}\right)\right\rangle _{V}\right],\label{eq:phi-rho-00}
\end{equation}
where 
\begin{equation}
C_{\phi}=3+\boldsymbol{P}_{s}\cdot\boldsymbol{P}_{\bar{s}}+c_{ii}^{(s\bar{s})}.
\end{equation}
In Eq. (\ref{eq:phi-rho-00}) the short-range average is taken inside
the $\phi$ meson with its spatial or momentum wave function.


\section{Inequalities for spin correlations}

The spin polarizations of hyperons and spin alignment of vector mesons
can be measured in experiments. From Eqs. (\ref{eq:av-baryons}) and
(\ref{eq:phi-rho-00}), we see that these observables depend on the
spin polarizations and correlation functions of constituent quarks
which have a huge number of unknown components or variables. In order
to identify the relevant features of quark polarizations and correlations,
we have to reduce the number of variables by making appropriate approximations.
In this section we will derive a few inequalities for spin correlations
implied by experimental data under these approximations. 

\subsection{Mean field approximation}

We make a mean-field approximation for all quark flavors in the short-range
averages in hyperon polarizations given by Eq. (\ref{eq:av-baryons}).
This means we do not distinguish flavor dependences in $\boldsymbol{P}_{q}$,
$c_{ij}^{(q_{1}q_{2})}$ and $c_{ijk}^{(q_{1}q_{2}q_{3})}$. By this
way we imply that $c_{ij}^{(q_{1}q_{2})}$ and $c_{ijk}^{(q_{1}q_{2}q_{3})}$
are symmetric for the interchange of any two spatial indices. There
are four types of terms: 
\begin{align}
B_{1}= & \left\langle c_{zi}^{(q_{1}q_{2})}P_{q_{3}}^{i}\right\rangle _{B}\equiv\left\langle c_{zi}^{(qq)}P_{q}^{i}\right\rangle _{B},\nonumber \\
B_{2}= & \left\langle \left(c_{ii}^{(q_{1}q_{2})}+\boldsymbol{P}_{q_{1}}\cdot\boldsymbol{P}_{q_{2}}\right)P_{q_{3}}^{z}\right\rangle _{B}\equiv\left\langle \left(c_{ii}^{(qq)}+\boldsymbol{P}_{q}\cdot\boldsymbol{P}_{q}\right)P_{q}^{z}\right\rangle _{B},\nonumber \\
B_{3}= & \left\langle c_{zii}^{(q_{1}q_{2}q_{3})}\right\rangle _{B}\equiv\left\langle c_{zii}^{(qqq)}\right\rangle _{B},\nonumber \\
B_{4}= & \left\langle c_{ii}^{(q_{1}q_{2})}+\boldsymbol{P}_{q_{1}}\cdot\boldsymbol{P}_{q_{2}}\right\rangle _{B}\left\langle P_{q_{3}}^{z}\right\rangle _{B}\equiv\left\langle c_{ii}^{(qq)}+\boldsymbol{P}_{q}\cdot\boldsymbol{P}_{q}\right\rangle _{B}\left\langle P_{q}^{z}\right\rangle _{B},
\end{align}
and we assume $B_{i}\ll1$ with $i=1,2,3,4$. Then we obtain from
Eq. (\ref{eq:av-baryons})
\begin{align}
P_{\varLambda}^{z}\approx & \left\langle P_{q}^{z}\right\rangle _{B}-2B_{1}-B_{2}-B_{3}+B_{4},\nonumber \\
P_{\Xi^{-}}^{z}\approx & P_{\varLambda}^{z},\nonumber \\
P_{\Omega}^{z}\approx & \frac{5}{3}\left\langle P_{q}^{z}\right\rangle _{B}+\frac{1}{3}\left(2B_{1}+B_{2}+B_{3}\right)-\frac{5}{3}B_{4}\nonumber \\
\approx & \frac{5}{3}P_{\varLambda}^{z}+2\left(2B_{1}+B_{2}+B_{3}\right)-\frac{10}{3}B_{4}.
\end{align}
Therefore, within the mean-field approximation we clearly see that
$P_{\Xi^{-}}\approx P_{\varLambda}$ and $P_{\Omega}\neq(5/3)P_{\varLambda}$
with 
\begin{equation}
P_{\Omega}^{z}-\frac{5}{3}P_{\varLambda}^{z}\approx2\left(2B_{1}+B_{2}+B_{3}\right)-\frac{10}{3}B_{4},
\end{equation}
which should be larger than zero in heavy-ion collisions at energies
lower than 20 GeV, as implied by data on $\Omega$ baryon's global
polarization in Au+Au collisions \citep{zhiwan_xu:qm2025}. This leads
to an inequality 
\begin{equation}
2B_{1}+B_{2}+B_{3}-\frac{5}{3}B_{4}>0,\label{eq:inequality-mean-f}
\end{equation}
when the collisional energy is less than 20 GeV in Au+Au collisions. 


The inequality (\ref{eq:inequality-mean-f}) clearly shows the presence
of the spin correlation in heavy-ion collisions at energies lower
than 20 GeV in Au+Au collisions. If quarks are not polarized the inequality
(\ref{eq:inequality-mean-f}) is reduced to $\left\langle c_{zii}^{(qqq)}\right\rangle _{B}>0$,
i.e. the three-quark correlation inside hyperons is non-vanishing.
On the other hand, if spin correlation functions are all zero, the
inequality (\ref{eq:inequality-mean-f}) is reduced to $\left\langle \left(\boldsymbol{P}_{q}\cdot\boldsymbol{P}_{q}\right)P_{q}^{z}\right\rangle _{B}>(5/3)\left\langle \boldsymbol{P}_{q}\cdot\boldsymbol{P}_{q}\right\rangle _{B}\left\langle P_{q}^{z}\right\rangle _{B}$,
i.e. the three quarks' polarizations are correlated inside hyperons. 


\subsection{With only $s$ quark correlations inside hyperons}

In this subsection we only consider short-range correlations among
strange quarks and antiquarks inside hyperons, neglecting mixed correlations
between quarks of different flavors. We can then relate the average
polarization of $\Xi$ and $\Omega$ hyperons with the spin alignment
of the $\phi$ meson. 


\subsubsection{Short- and long-range components of spin polarizations and correlations}

The spin polarizations for quarks and antiquarks $\boldsymbol{P}_{q/\bar{q}}$
can be decomposed into the long and short range components. The long
range component $\overline{\boldsymbol{P}}_{q/\bar{q}}$ is induced
by local vorticity and has the same sign for the quark and antiquark.
The short range component $\widetilde{\boldsymbol{P}}_{q/\bar{q}}$
is from vector fields and has the opposite sign for the quark and
antiquark. Namely the polarization vectors can be written as 
\begin{align}
\boldsymbol{P}_{q}= & \overline{\boldsymbol{P}}_{q}+\widetilde{\boldsymbol{P}}_{q},\nonumber \\
\boldsymbol{P}_{\bar{q}}= & \overline{\boldsymbol{P}}_{\bar{q}}-\widetilde{\boldsymbol{P}}_{\bar{q}},\label{eq:two-comp}
\end{align}
for $q=u,d,s$. We assume that the averages of short range components
inside hyperons are restricted by their wave functions and satisfy
\begin{align}
\left\langle \widetilde{\boldsymbol{P}}_{q}\right\rangle _{B}= & 0,\nonumber \\
\left\langle \left(\widetilde{\boldsymbol{P}}_{q_{1}}\cdot\widetilde{\boldsymbol{P}}_{q_{2}}\right)\widetilde{\boldsymbol{P}}_{q_{3}}\right\rangle _{B}= & 0,\nonumber \\
\left\langle \widetilde{\boldsymbol{P}}_{q}\cdot\widetilde{\boldsymbol{P}}_{q}\right\rangle _{B}\neq & 0,\label{eq:polar-fluctuation}
\end{align}
where $B$ represents the baryon wave function. So we have $\left\langle \boldsymbol{P}_{q}\right\rangle _{B}=\overline{\boldsymbol{P}}_{q}$.
We can further assume that long range components do not distinguish
the quark, antiquark and flavor, $\overline{\boldsymbol{P}}_{q_{i}}=\overline{\boldsymbol{P}}_{\bar{q}_{j}}$,
with $i,j=u,d,s$. All spin correlation matrices $c_{ij}^{(q_{1}q_{2})}$
and $c_{ijk}^{(q_{1}q_{2}q_{3})}$ are assumed to be of short range. 

When taking the short range average of $\delta\rho_{B}$ over the
hadron volume we can approximate 
\begin{align}
\left\langle c_{ij}^{(q_{1}q_{2})}P_{q_{3}}^{k}\right\rangle _{B}\approx & \left\langle c_{ij}^{(q_{1}q_{2})}\right\rangle _{B}\overline{P}_{q_{3}}^{k},\nonumber \\
\left\langle \left(\boldsymbol{P}_{q_{1}}\cdot\boldsymbol{P}_{q_{2}}\right)P_{q_{3}}^{z}\right\rangle _{B}\approx & \left\langle \widetilde{\boldsymbol{P}}_{q_{1}}\cdot\widetilde{\boldsymbol{P}}_{q_{2}}\right\rangle _{B}\overline{P}_{q_{3}}^{z}+\left\langle \widetilde{\boldsymbol{P}}_{q_{1}}^{i}\widetilde{\boldsymbol{P}}_{q_{3}}^{z}\right\rangle _{B}\overline{P}_{q_{2}}^{i},\nonumber \\
 & +\left\langle \widetilde{\boldsymbol{P}}_{q_{2}}^{i}\widetilde{\boldsymbol{P}}_{q_{3}}^{z}\right\rangle _{B}\overline{P}_{q_{1}}^{i},\nonumber \\
\overline{\boldsymbol{P}}_{q_{1}}\cdot\overline{\boldsymbol{P}}_{q_{2}}\ll & \left\langle \widetilde{\boldsymbol{P}}_{q_{1}}\cdot\widetilde{\boldsymbol{P}}_{q_{2}}\right\rangle _{B}\ll1.\label{eq:rules-short-av}
\end{align}
In this way we can relate the average polarization of $\Xi$ and $\Omega$
hyperons with the spin alignment of the $\phi$ meson. 


\subsubsection{Spin polarizations of hyperons and spin alignment of $\phi$ meson}

With the decomposition (\ref{eq:two-comp}) and the approximations
in short range averages in Eqs. (\ref{eq:polar-fluctuation}) and
(\ref{eq:rules-short-av}), we obtain 
\begin{align}
P_{\Lambda}^{z}\approx & \overline{P}_{s}^{z}=\overline{P}_{q}^{z},\nonumber \\
P_{\Xi^{-}}^{z}\approx & \overline{P}_{q}^{z}\left(1+\epsilon_{\Xi}\right)-\frac{4}{3}\delta,\nonumber \\
P_{\Omega}^{z}\approx & \frac{5}{3}\overline{P}_{q}^{z}\left(1+\epsilon_{\Omega}\right)+\frac{2}{3}\delta+\frac{1}{3}\delta^{\prime},\label{eq:results-hyperons}
\end{align}
where we have defined 
\begin{align}
\epsilon_{\Xi}\equiv & -\frac{4}{3}\left\langle \left(\widetilde{P}_{s}^{z}\right)^{2}+c_{zz}^{(ss)}\right\rangle _{B},\nonumber \\
\epsilon_{\Omega}\equiv & -\frac{4}{5}\left\langle \widetilde{\boldsymbol{P}}_{s}\cdot\widetilde{\boldsymbol{P}}_{s}+c_{ii}^{(ss)}\right\rangle _{B}+\frac{2}{5}\left\langle \left(P_{s}^{z}\right)^{2}+c_{zz}^{(ss)}\right\rangle _{B},\nonumber \\
\delta\equiv & \left\langle c_{zx}^{(ss)}\right\rangle _{B}\overline{P}_{q}^{x}+\left\langle c_{zy}^{(ss)}\right\rangle _{B}\overline{P}_{q}^{y},\nonumber \\
\delta^{\prime}\equiv & \left\langle c_{zii}^{(sss)}\right\rangle _{B}.\label{eq:results-hyperons-def}
\end{align}
In the derivation of Eq. (\ref{eq:results-hyperons}) we have kept
$c^{(ss)}$ and $c^{(sss)}$ only and neglected all other spin correlation
functions.


For the spin alignment of the $\phi$ meson, we obtain from Eq. (\ref{eq:phi-rho-00})
\begin{equation}
\rho_{00}^{\phi}\approx\frac{1}{3}\left[1+\frac{2}{3}\left\langle \widetilde{\boldsymbol{P}}_{s}\cdot\widetilde{\boldsymbol{P}}_{\bar{s}}+c_{ii}^{(s\bar{s})}\right\rangle _{V}-2\left\langle \widetilde{P}_{s}^{z}\widetilde{P}_{\bar{s}}^{z}+c_{zz}^{(s\bar{s})}\right\rangle _{V}\right].\label{eq:rho00-phi}
\end{equation}
If the short range spin polarization and correlation are caused by
vector fields which distinguish particles and antiparticles, we can
make an approximation on the short-range effective correlations 
\begin{align}
\left\langle \widetilde{\boldsymbol{P}}_{s}\cdot\widetilde{\boldsymbol{P}}_{\bar{s}}+c_{ii}^{(s\bar{s})}\right\rangle _{V}\approx & -\eta\left\langle \widetilde{\boldsymbol{P}}_{s}\cdot\widetilde{\boldsymbol{P}}_{s}+c_{ii}^{(ss)}\right\rangle _{B},\nonumber \\
\left\langle \widetilde{P}_{s}^{z}\widetilde{P}_{\bar{s}}^{z}+c_{zz}^{(s\bar{s})}\right\rangle _{V}\approx & -\eta\left\langle \widetilde{P}_{s}^{z}\widetilde{P}_{s}^{z}+c_{zz}^{(ss)}\right\rangle _{B},\label{eq:phi-baryon-rel}
\end{align}
where $\eta>0$ is a proportionality constant. Then Eq. (\ref{eq:rho00-phi})
becomes 
\begin{equation}
\rho_{00}^{\phi}\approx\frac{1}{3}\left(1+\epsilon_{\phi}\right),\label{eq:rho00-phi-1}
\end{equation}
where we defined 
\begin{equation}
\epsilon_{\phi}\equiv-\frac{2}{3}\eta\left\langle \widetilde{\boldsymbol{P}}_{s}\cdot\widetilde{\boldsymbol{P}}_{s}+c_{ii}^{(ss)}\right\rangle _{B}+2\eta\left\langle \widetilde{P}_{s}^{z}\widetilde{P}_{s}^{z}+c_{zz}^{(ss)}\right\rangle _{B}.
\end{equation}

The quantities $\epsilon_{\Xi}$, $\epsilon_{\Omega}$ and $\epsilon_{\phi}$
depend on spin correlations in the spin quantization direction and
in transverse directions perpendicular to it. Therefore, we can define
longitudinal and transverse effective correlations as 
\begin{align}
A_{L}= & \left\langle \left(\widetilde{P}_{s}^{z}\right)^{2}+c_{zz}^{(ss)}\right\rangle _{B},\nonumber \\
A_{T}= & \left\langle \left(\widetilde{P}_{s}^{x}\right)^{2}+\left(\widetilde{P}_{s}^{y}\right)^{2}+c_{xx}^{(ss)}+c_{yy}^{(ss)}\right\rangle _{B},\label{eq:def-c12}
\end{align}
in terms of which the $\epsilon$ quantities in Eqs. (\ref{eq:results-hyperons-def})
and (\ref{eq:rho00-phi-1}) can be rewritten as 
\begin{align}
\epsilon_{\Xi}= & -\frac{4}{3}A_{L},\nonumber \\
\epsilon_{\Omega}= & -\frac{2}{5}(2A_{T}+A_{L}),\nonumber \\
\epsilon_{\phi}= & -\frac{2}{3}\eta(A_{T}-2A_{L}).\label{eq:deviation}
\end{align}


\subsubsection{Inequalities for quark spin correlations}

By comparing Eqs. (\ref{eq:results-hyperons}) and (\ref{eq:rho00-phi-1})
with the experimental data, we obtain the following constraints:

\begin{align}
P_{\Xi^{-}}-P_{\Lambda}\approx & -\frac{4}{3}\overline{P}_{q}^{z}A_{L}-\frac{4}{3}\delta\approx0,\nonumber \\
P_{\Omega}-\frac{5}{3}P_{\Lambda}\approx & -\frac{2}{3}\overline{P}_{q}^{z}(2A_{T}+A_{L})+\frac{2}{3}\delta+\frac{1}{3}\delta^{\prime}>0,\nonumber \\
\rho_{00}^{\phi}-\frac{1}{3}= & \frac{2}{9}\eta(2A_{L}-A_{T})>0,\label{eq:exp_constraints}
\end{align}
which leads to one equality and two inequalities for quark spin correlations
\begin{align}
\delta\approx-\overline{P}_{q}^{z}A_{L} & ,\nonumber \\
\delta^{\prime}-4\overline{P}_{q}^{z}(A_{T}+A_{L})> & 0,\nonumber \\
2A_{L}-A_{T}> & 0.\label{eq:ineq-a}
\end{align}
The explicit expressions of above equality and inequalities are 
\begin{align}
\left\langle c_{zx}^{(ss)}\right\rangle _{B}\overline{P}_{q}^{x}+\left\langle c_{zy}^{(ss)}\right\rangle _{B}\overline{P}_{q}^{y}\approx & -\overline{P}_{q}^{z}\left\langle \left(\widetilde{P}_{s}^{z}\right)^{2}+c_{zz}^{(ss)}\right\rangle _{B},\nonumber \\
4\overline{P}_{q}^{z}\left\langle \widetilde{\boldsymbol{P}}_{s}\cdot\widetilde{\boldsymbol{P}}_{s}+c_{ii}^{(ss)}\right\rangle _{B}< & \left\langle c_{zii}^{(sss)}\right\rangle _{B},\nonumber \\
\left\langle \widetilde{\boldsymbol{P}}_{s}\cdot\widetilde{\boldsymbol{P}}_{s}+c_{ii}^{(ss)}\right\rangle _{B}< & 3\left\langle \left(\widetilde{P}_{s}^{z}\right)^{2}+c_{zz}^{(ss)}\right\rangle _{B}.\label{eq:case2-ineq}
\end{align}
The results in (\ref{eq:case2-ineq}) clearly show the presence of
spin correlations between $s$ quarks in hyperons. If we neglect the
three-quark spin correlation, i.e. $\left\langle c_{zii}^{(sss)}\right\rangle _{B}=0$,
from the first inequality we obtain non-vanishing spin correlation
between two $s$ quarks 
\begin{equation}
c_{xx}^{(ss)}+c_{yy}^{(ss)}+c_{zz}^{(ss)}<0.
\end{equation}
The second inequality in (\ref{eq:case2-ineq}) looks like a quadrupole
structure in spin correlations.


\subsubsection{Estimates of quark spin correlations}

In order to extract some quantitative estimates of quark spin correlations
inside hyperons, in this subsection we make a further assumption that
long-range correlations are only along the $z$ direction, which is
assumed to be the direction of the global orbital angular momentum
in non-central heavy-ion collisions. So we have 
\begin{equation}
\overline{P}_{q}^{z}\approx P_{\Lambda},\;\overline{P}_{q}^{x}\approx\overline{P}_{q}^{y}\approx0.
\end{equation}
This leads to $\delta\approx0$ from the first equation of (\ref{eq:exp_constraints}).
Then, the experimental constraints on $P_{\Xi^{-}}$, $P_{\Omega}$
and $\rho_{00}^{\phi}$ in Eq. (\ref{eq:exp_constraints}) become:
\begin{align}
P_{\Xi^{-}}-P_{\Lambda} & \approx-\frac{4}{3}P_{\Lambda}A_{L}\approx0,\nonumber \\
P_{\Omega}-\frac{5}{3}P_{\Lambda} & \approx-\frac{2}{3}P_{\Lambda}(2A_{T}+A_{L})+\frac{1}{3}\delta^{\prime}>0,\nonumber \\
\rho_{00}^{\phi}-\frac{1}{3} & \approx\frac{2}{9}\eta(2A_{L}-A_{T}).\label{eq:delta0}
\end{align}
This is a system of three equations in four unknown variables: $A_{L}$,
$A_{T}$, $\delta^{\prime}$ and $\eta$. From the first equation
we can extract $A_{L}$: the fact that there is no significant difference
between $\Lambda$ and $\Xi$ global polarization implies that $A_{L}\approx0$,
i.e., longitudinal effective correlations between $s$ quarks are
negligible. We can consider $\eta$ as a fixed parameter, so that
the other two variables, $A_{T}$ and $\delta^{\prime}$, can be determined
from the second and third equations in (\ref{eq:delta0}) 
\begin{align}
A_{T} & \approx\frac{3}{2\eta}\left(1-3\rho_{00}^{\phi}\right),\nonumber \\
\delta^{\prime} & \approx3P_{\Omega}-P_{\Lambda}\left[5-\frac{6}{\eta}\left(1-3\bar{\rho}_{00}^{\phi}\right)\right].\label{eq:delta_prime}
\end{align}

\begin{figure}
\centering \includegraphics[width=0.45\textwidth]{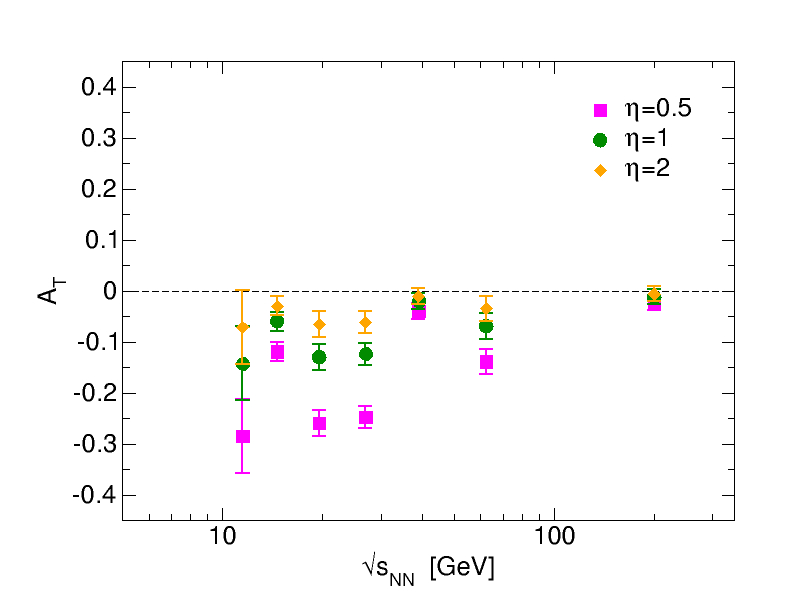}
\includegraphics[width=0.45\textwidth]{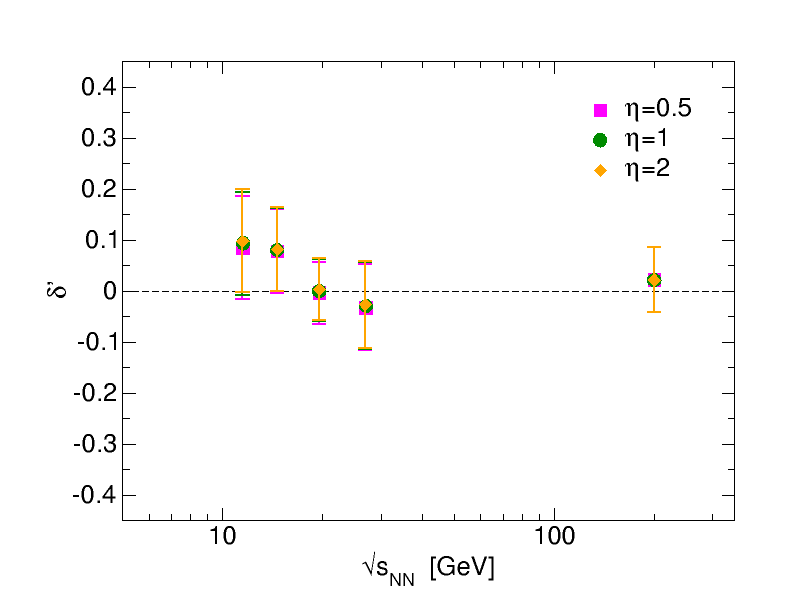}
\caption{Estimate of two-quark transverse correlations $A_{T}$ (left panel)
and the three-quark correlations $\delta^{\prime}$ (right panel)
as a function of collision energy for $\eta=0.5,1,2$.}
\label{fig:A_T-delta_prime}
\end{figure}

We estimate the transverse two-particle correlations $A_{T}$ and
the three-particle correlations $\delta^{\prime}$ as a function of
collision energy from available experimental data on $P_{\Lambda}$,
$P_{\Omega}$ and $\rho_{00}^{\phi}$. We also estimate the uncertainties
on $A_{T}$ and $\delta^{\prime}$ in the extraction, by considering
the experimental errors on $P_{\Omega}$ and $\rho_{00}^{\phi}$,
while neglecting those on $P_{\Lambda}$ (being them small). In Fig.
\ref{fig:A_T-delta_prime} we show results for the two-particle transverse
correlations $A_{T}$ (left panel) and the three-quark correlations
$\delta^{\prime}$ (right panel) as a function of collision energy
for different values of the parameter $\eta$. For the extraction
of $A_{T}$ we consider all the collision energies where there are
data on $\rho_{00}^{\phi}$: 11.5, 14.6, 19.6, 27.0, 39.0, 62.4, 200
GeV. Regarding $\delta^{\prime}$, we can extract it only at the common
collision energies of the different spin observable measurements:
11.5, 14.6, 19.6, 27.0, 200 GeV. From Fig. \ref{fig:A_T-delta_prime}
we see that transverse effective correlation are negative: 
\begin{equation}
\left\langle \left(\widetilde{P}_{s}^{x}\right)^{2}+\left(\widetilde{P}_{s}^{y}\right)^{2}+c_{xx}^{(ss)}+c_{yy}^{(ss)}\right\rangle _{B}<0.
\end{equation}
This implies that transverse genuine correlations between strange
quarks are non-vanishing and negative, i.e., $\left\langle c_{xx}^{(ss)}+c_{yy}^{(ss)}\right\rangle _{B}<0$.

\section{Summary and conclusion}

The global spin polarization of hyperons have been investigated by
including spin correlation effects among their constituent quarks.
Based on density operators for multi-quarks (including antiquarks)
in spin-momentum space, we derived the Spin Wigner Functions (SWF)
for vector mesons or baryons by computing elements of density operators
between two spin-momentum states of the vector meson or baryon. The
SWF are actually spin density matrices in phase space for vector mesons
and baryons which are expressed as phase space integrals of their
constituent quarks' reduced spin density matrices weighted by their
coordinate wave functions, constituent quark distributions and CG
coefficients. The short range correlation is defined by the average
over phase space of quarks (antiquarks) inside the hadron, while the
long range correlation is by the average over phase space of the hadron.
The available data in global spin polarizations of hyperons and spin
alignments of vector mesons provide constraints on phase space functions
of the spin polarization and correlation appearing in the reduced
spin density matrices for constituent quarks, which consist of large
number of unknown variables. In order to reduce the number of variables
and derive meaningful and simplified constraints, we make two approximations.
In the first one, which is called the mean field approximation, we
do not distinguish different flavors in $\boldsymbol{P}_{q}$, $c_{ij}^{(q_{1}q_{2})}$
and $c_{ijk}^{(q_{1}q_{2}q_{3})}$. In the second approximation, we
only consider spin correlation among $s$ quarks and neglect others.
In each approximation, we derived a set of inequalities for spin polarization
and correlation functions. These inequalities might provide possible
clues for the presence of spin correlation inside hyperons at lower
collision energies.
\begin{acknowledgments}
We thank V. Greco, Z.-T. Liang and D. Rischke for insightful discussons.
Q.W. is supported in part by the National Natural Science Foundation
of China (NSFC) under Grant No. 12135011. L.O. acknowledges support
from the European Union -- Next Generation EU, Mission 4, Component
2, Investment line 1.2, CUP E63C22002960006, for the project HEFESTUS.
\end{acknowledgments}

\bibliographystyle{unsrt}
\bibliography{ref-correlation-hyperon}

\end{document}